\documentclass[]{spie}  

\usepackage{amsmath,amsfonts,amssymb}
\usepackage{graphicx}
\usepackage[colorlinks=true, allcolors=blue]{hyperref}

\usepackage{multicol}

\usepackage[bottom=0.6in,left=0.6in,right=0.6in]{geometry}

\def\PRL{{\em Phys. Rev. Lett.}}
\def\PRD{{\em Phys. Rev.} D}

\def\aap{{\em A\&A}} 
\def\APJ{{\em ApJ}} 

\def\Ad{A_\mathrm{d}}

\def\As{A_\mathrm{sync}}

\def\ad{\alpha_\mathrm{d}}
\def\as{\alpha_\mathrm{s}}

\def\ddp{\Delta_\mathrm{d}}

\def\dasi{\textsc{Dasi}}
\def\bicep{\textsc{Bicep}}
\def\bicepone{\textsc{Bicep1}}
\def\biceptwo{\textsc{Bicep2}}
\def\bicepthree{\textsc{Bicep3}}
\def\biceptng{\textsc{Bicep} Array} 
\def\keck{{\it Keck}}
\def\keckarray{{\it Keck Array}}
\def\bk{{\bicep/\keck}}
\def\planck{{\it Planck}} 
\def\spider{{Spider}} 
\def\quiet{\textsc{Quiet}}
\def\spt{\textsc{SPT}}

\def\sptpol{\textsc{SPTpol}}
\def\sptthreeg{\textsc{SPT-3G}}

\def\abs{\textsc{Abs}}
\def\act{\textsc{ACT}}

\def\polarbear{\textsc{Polarbear}}
\def\wmap{\textsc{WMAP}}
\def\cmbsfour{{CMB-S4}}
\def\be{\begin{equation}}
\def\ee{\end{equation}}
\def\bea{\begin{eqnarray}}
\def\eea{\end{eqnarray}}

\title{Measurements of Degree-Scale B-mode Polarization with the  BICEP/Keck Experiments at South Pole}
\author[a]{The \bk\ collaboration: P.~A.~R.~Ade}
\author[b]{Z.~Ahmed}
\author[c]{R.~W.~Aikin}
\author[d]{K.~D.~Alexander}
\author[d]{D.~Barkats}
\author[e]{S.~J.~Benton}
\author[f]{C.~A.~Bischoff}
\author[c,g]{J.~J.~Bock}
\author[d]{H.~Boenish}
\author[d]{R.~Bowens-Rubin}
\author[c]{J.~A.~Brevik}
\author[d]{I.~Buder}
\author[h]{E.~Bullock}
\author[d,i]{V.~Buza}
\author[d]{J.~Connors}
\author[d]{J.~Cornelison}
\author[g]{B.~P.~Crill}
\author[j]{M.~Crumrine}
\author[d]{M.~Dierickx}
\author[k]{L.~Duband}
\author[i]{C.~Dvorkin}
\author[l,m]{J.~P.~Filippini}
\author[j]{S.~Fliescher}
\author[n]{J.~Grayson}
\author[j]{G.~Hall}
\author[o]{M.~Halpern}
\author[d]{S.~Harrison}
\author[c,g]{S.~R.~Hildebrandt}
\author[p]{G.~C.~Hilton}
\author[c]{H.~Hui}
\author[n,c,p]{K.~D.~Irwin}
\author[n]{J. Kang}
\author[d,q]{K.~S.~Karkare}
\author[n]{E.~Karpel}
\author[r]{J.~P.~Kaufman}
\author[r]{B.~G.~Keating}
\author[c]{S.~Kefeli}
\author[n]{S.~A.~Kernasovskiy}
\author[d,i]{J.~M.~Kovac}
\author[n,b]{C.~L.~Kuo}
\author[j]{K.~Lau}
\author[q]{N.~A.~Larsen}
\author[q]{E.~M.~Leitch}
\author[c]{M.~Lueker}
\author[g]{K.~G.~Megerian}
\author[c]{L.~Moncelsi}
\author[s]{T.~Namikawa}
\author[e,t]{B.~Netterfield}
\author[g]{H.~T.~Nguyen}
\author[c,g]{R.~O'Brient}
\author[n,b]{R.~W.~Ogburn~IV}
\author[f]{S.~Palladino}
\author[h,j]{C.~Pryke}
\author[d,*]{B.~Racine}
\author[d]{S.~Richter}
\author[j]{R.~Schwarz}
\author[c]{A.~Schillaci}
\author[q,u]{C.~D.~Sheehy}
\author[c]{A.~Soliman}
\author[d]{T.~St.~Germaine}
\author[c,g]{Z.~K.~Staniszewski}
\author[c]{B.~Steinbach}
\author[a]{R.~V.~Sudiwala}
\author[c,r]{G.~P.~Teply}
\author[n,b]{K.~L.~Thompson}
\author[n]{J.~E.~Tolan}
\author[a]{C.~Tucker}
\author[g]{A.~D.~Turner}
\author[f]{C.~Umilt\`{a}}
\author[q,v]{A.~G.~Vieregg}
\author[c]{A.~Wandui}
\author[g]{A.~C.~Weber}
\author[o]{D.~V.~Wiebe}
\author[j]{J.~Willmert}
\author[d,i]{C.~L.~Wong}
\author[n,q]{W.~L.~K.~Wu}
\author[n]{H.~Yang}
\author[n,b]{K.~W.~Yoon}
\author[c]{C.~Zhang}

\affil[a]{School of Physics and Astronomy, Cardiff University, Cardiff, CF24 3AA, United Kingdom}
\affil[b]{Kavli Institute for Particle Astrophysics and Cosmology, SLAC, Menlo Park, CA 94025, USA}
\affil[c]{Department of Physics, California Institute of Technology, Pasadena, CA 91125, USA}
\affil[d]{Harvard-Smithsonian Center for Astrophysics, Cambridge, MA 02138, USA}
\affil[e]{Department of Physics, University of Toronto, Toronto, Ontario, M5S 1A7, Canada}
\affil[f]{Department of Physics, University of Cincinnati, Cincinnati, OH 45221, USA}
\affil[g]{Jet Propulsion Laboratory, Pasadena, CA 91109, USA}
\affil[h]{Minnesota Institute for Astrophysics, University of Minnesota, Minneapolis, MN 55455, USA}
\affil[i]{Department of Physics, Harvard University, Cambridge, MA 02138, USA}
\affil[j]{School of Physics and Astronomy, University of Minnesota, Minneapolis, MN 55455, USA}
\affil[k]{Service des Basses Temp´eratures, Commissariat `a l’Energie Atomique, 38054 Grenoble, France}
\affil[l]{Department of Physics, University of Illinois at Urbana-Champaign, Urbana, IL 61801, USA}
\affil[m]{Department of Astronomy, University of Illinois at Urbana-Champaign, Urbana, IL 61801, USA}
\affil[n]{Department of Physics, Stanford University, Stanford, CA 94305, USA}

\affil[o]{Department of Physics and Astronomy, University of British Columbia,Vancouver, V6T 1Z1, Canada}
\affil[p]{National Institute of Standards and Technology, Boulder, CO 80305, USA}
\affil[q]{Kavli Institute for Cosmological Physics, University of Chicago, Chicago, IL 60637, USA}
\affil[r]{Department of Physics, University of California at San Diego, La Jolla, CA 92093, USA}
\affil[s]{Leung Center for Cosmology \& Particle Astrophysics, National Taiwan University, Taipei 10617, TW}
\affil[t]{Canadian Institute for Advanced Research, Toronto, Ontario, M5G 1Z8, Canada}
\affil[u]{Physics Department, Brookhaven National Laboratory, Upton, NY 11973}
\affil[v]{Department of Physics, Enrico Fermi Institute, University of Chicago, Chicago, IL 60637}

\authorinfo{*Corresponding author: B.Racine, \href{mailto:benjamin.racine@cfa.harvard.edu}{benjamin.racine@cfa.harvard.edu}}
\begin{document} 
\maketitle

\begin{abstract}
The \bicep\ and \keckarray\ experiments are a suite of small-aperture refracting telescopes observing the microwave sky from the South Pole. They target the degree-scale $B$-mode polarization signal imprinted in the Cosmic Microwave Background (CMB) by primordial gravitational waves. Such a measurement would shed light on the physics of the very early universe. While \biceptwo\ observed for the first time a $B$-mode signal at 150 GHz, higher frequencies from the \planck\ satellite showed that it could be entirely due to the polarized emission from Galactic dust, though uncertainty remained high. \keckarray\ has been observing the same region of the sky for several years, with an increased detector count, producing the deepest polarized CMB maps to date. New detectors at 95 GHz were installed in 2014, and at 220 GHz in 2015. These observations enable a better constraint of galactic foreground emissions, as presented here.
In 2015, \biceptwo\ was replaced by \bicepthree, a 10 times higher throughput telescope observing at 95 GHz, while \keckarray\  is now focusing on higher frequencies. In the near future, \biceptng\ will replace \keckarray, and will allow unprecedented sensitivity to the gravitational wave signal. High resolution observations from the South Pole Telescope (\spt) will also be used to remove the lensing contribution to $B$-modes.
\end{abstract}

\section{Introduction}

Our standard model of Cosmology, $\Lambda$CDM,  is able to statistically describe our observable universe with only six cosmological parameters. We know these to percent-level precision, in large parts from CMB data\cite{2016A&A...594A..13P}. As it has been extensively presented at the 2018 Moriond Conference, while other astrophysical probes are in agreement with this model, we still don't understand its main components, the so-called dark energy and dark matter. Many theoretical and experimental studies are ongoing to study them or find alternative models.

\begin{figure}[!t]
\centerline{\includegraphics[trim={0 1cm 0 4cm},width=0.75\linewidth]{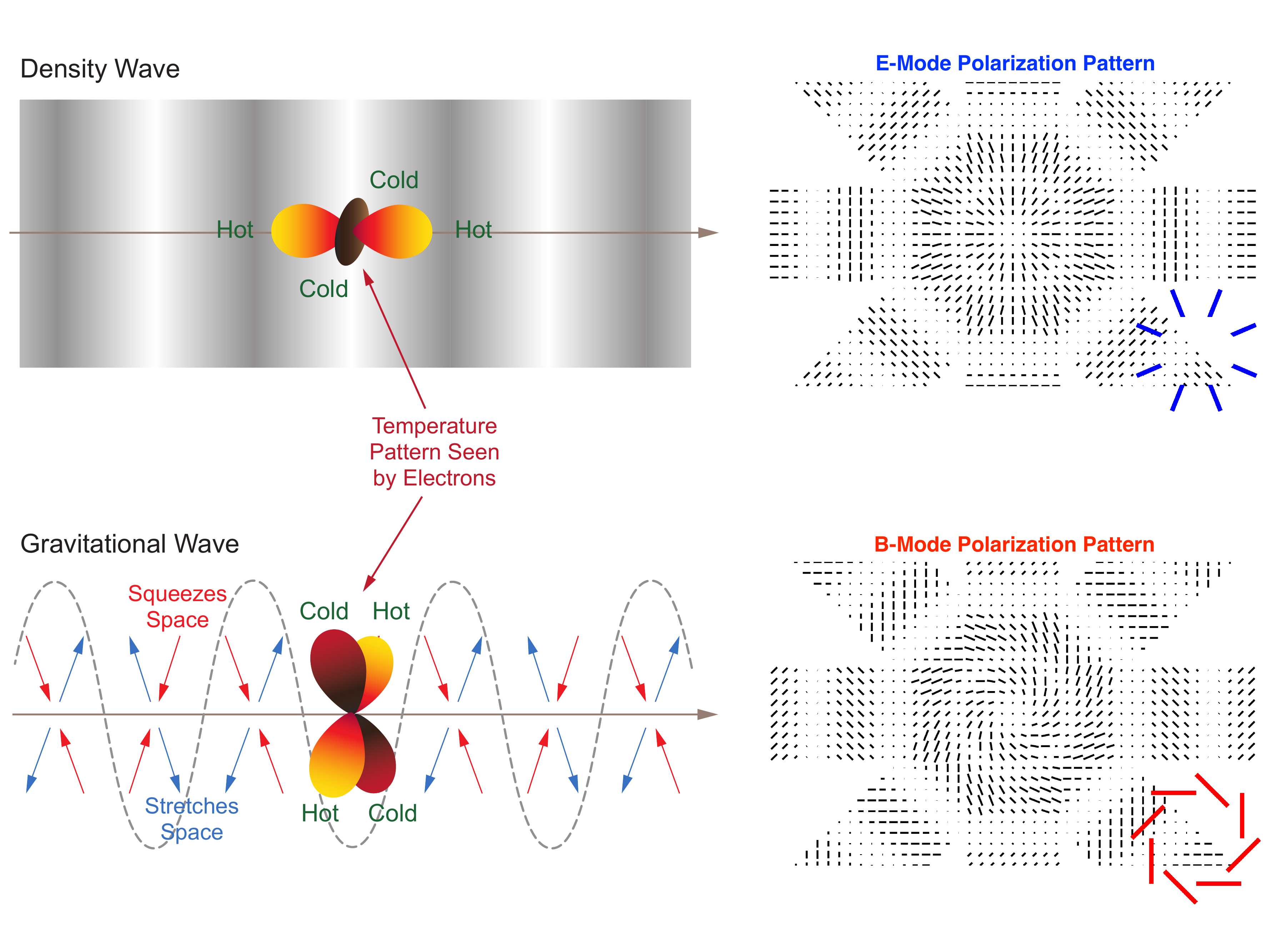}}
\caption[]{Illustration of CMB polarization generation. On the top left, we show a density wave on the last scattering surface (LSS), with peaks in dark gray and troughs in white. In the reference frame of the electron in a cold spot, there is a quadrupolar inflow of photons; through Thomson scattering, the photons traveling perpendicular to the plane will have a linear polarization aligned with the cold stream direction. On the upper right, we see how the polarization pattern looks like for three crossing plane waves. These curl-free pattern are called $E$-modes. On the lower left, we show the signal from a x-polarized gravitational wave traveling across the LSS, (if exactly along the LSS, it cancels, here the gray arrow shows a projected wave vector). On the lower right, we show the gradient-free $B$-mode patterns. Note that a +-polarized gravitational wave produces $E$-mode patterns.}
\label{fig:HighresEB}
\end{figure}

Another perhaps greater mystery resides in the very early history of our Universe. The leading paradigm, Inflation, states that there was a period of exponential expansion, $\sim 10^{-35}$s after the Big Bang. This extreme phenomenon could explain the homogeneity and the flatness of our Universe\cite{1981PhRvD..23..347G} and would naturally generate and drive quantum fluctuations to cosmological scales\cite{{1981ZhPmR..33..549M}}. These density perturbations are the seeds of temperature anisotropies of the CMB, as well as all structures of our Universe. Thomson scattering at the last scattering surface also generates linear polarization of some of these photons (see $E$-modes in Figure \ref{fig:HighresEB}).

CMB measurements allow direct testing of predictions from inflation: namely that these initial perturbations are Gaussian, adiabatic, almost but not quite scale-invariant, and that our observable patch has a flat geometry. The coherence of these fluctuations at scales larger than the causal horizon at recombination, mainly visible in the anti-correlation of temperature and $E$-mode polarization, is also a prediction, which has ruled out other structure formation mechanisms, such as topological defects.

Inflation is also believed to have perturbed the metric of space-time itself, generating primordial gravitational waves (PGW). This is not the case of most alternative theories, and such a detection would be considered a major observational anchor of inflationary theories. As can be seen in Figure \ref{fig:HighresEB}, PGWs produce a curl-component polarization pattern in the CMB, called $B$-modes\cite{1997EBSeljak,1997EBKamionkowski}. We often characterize the amplitude of this signal at the power spectrum level, using the tensor-to-scalar ratio $r$. 

\begin{figure*}[!h]
\begin{minipage}{0.57\textwidth}
\resizebox{\textwidth}{!}{\includegraphics[trim={0 0.cm 0 0.cm}]{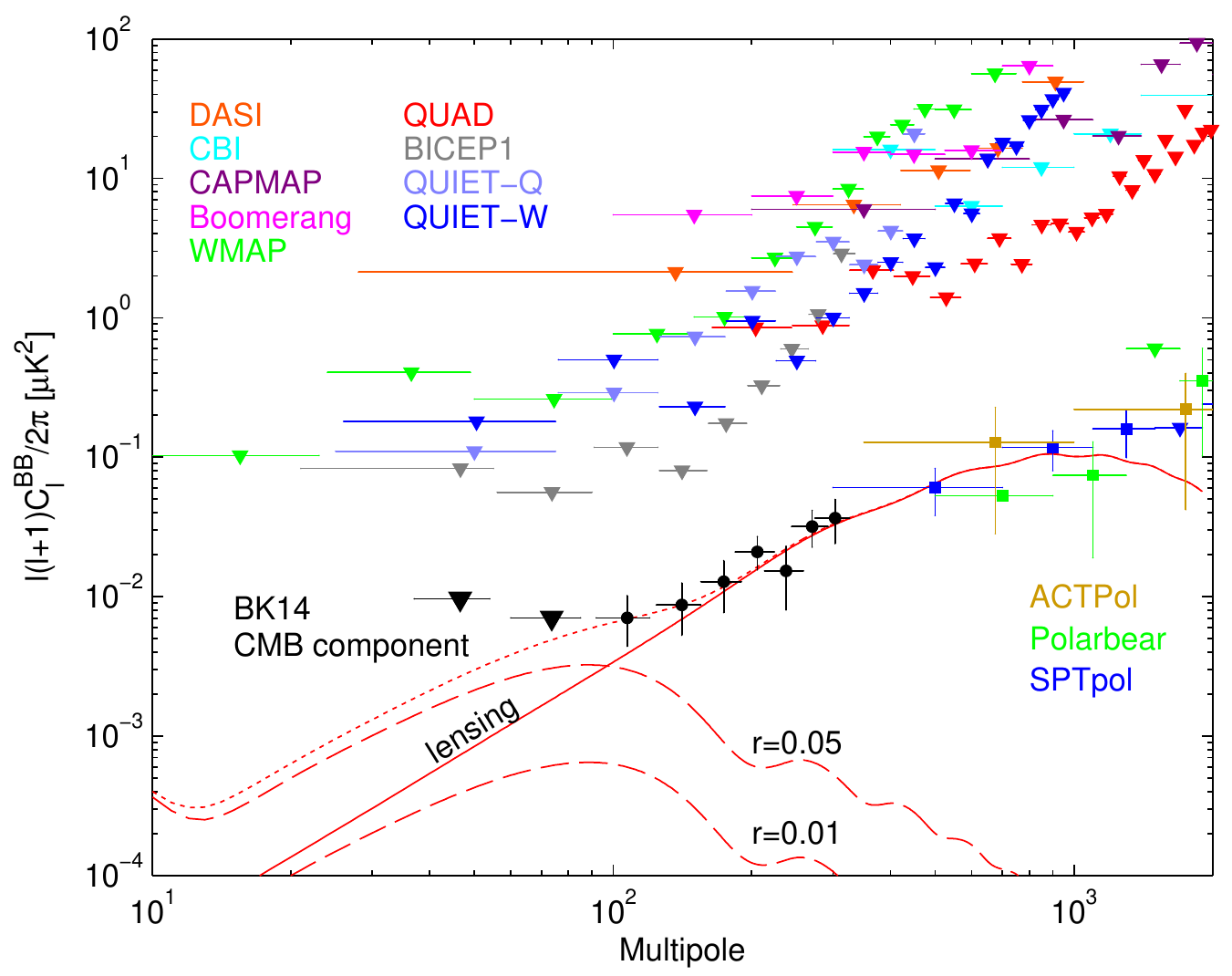}}
\end{minipage}
\begin{minipage}{0.5\textwidth}
\begin{center}
\scalebox{0.85}{
\begin{tabular}[b]{|l|l|l|}
\hline
               & arXiv      & $\sigma(r)$ \\
\hline
\dasi\         & \href{https://arxiv.org/abs/astro-ph/0409357}{0409357}    & 7.5 \\
\bicepone\ 2yr & \href{https://arxiv.org/abs/0906.1181}{0906.1181}  & 0.28 \\
\wmap\ 7yr     & \href{https://arxiv.org/abs/1001.4538}{1001.4538}  & 1.1 \\
\quiet\-Q       & \href{https://arxiv.org/abs/1012.3191}{1012.3191}  & 0.97 \\
\quiet\-W       & \href{https://arxiv.org/abs/1207.5034}{1207.5034}  & 0.85 \\
\bicepone\ 3yr & \href{https://arxiv.org/abs/1310.1422}{1310.1422}  & 0.25 \\
\biceptwo\     & \href{https://arxiv.org/abs/1403.3985}{1403.3985}  & 0.10 \\
BK13/\planck\ & \href{https://arxiv.org/abs/1502.00612}{1502.00612} & 0.034 \\
BK14/W/P        & \href{https://arxiv.org/abs/1510.09217}{1510.09217} & 0.024 \\
\abs\       & \href{https://arxiv.org/abs/1801.01218}{1801.01218} & 0.7 \\ 
\hline
\end{tabular}
}
\end{center}
\end{minipage}
\caption{\small {\it Left:} Published $B$-modes power spectra as of March 2018. The red lines show the theoretical predictions for the GW signal (dashed) for $r=0.05$ and $r=0.01$, peaking at degree scales  ($\ell \sim 80$), as well as the lensing signal (solid), peaking at arcminute scales ($\ell \sim 1000$). The dots show detected signals, and triangles are 95\% upper limits.  The black points show the $B$-mode
 spectrum from \bk\ data (up to 2014) with \wmap\ and \planck\ after removing galactic foreground contamination, with a $> 8\sigma$ detection of gravitational lensing. Other data points from telescopes focusing on the small scale lensing signal are shown. \small {\it Right:} Compilation of the published sensitivity to $r$. 
}
\label{fig:bk_vs_world}
\end{figure*}

In 2015, \planck\ placed the tightest constraints on $r$ achievable with CMB temperature data only ($r_{0.002}<0.11$)\cite{2016A&A...594A..13P}. Progress can now come from polarized data. The best constraints as of March 2018 come from combining  \bk\ data with \planck\ and \wmap\ data (described in Section \ref{sec:BK}) : $r_{0.05} < 0.07$ at 95\% confidence. 

While primordial $B$-modes have a clear signature in the polarized sky, other sources can mimic such a signal. Along their trajectory since the last scattering surface, photons get deviated by the gravitational potentials of massive structures in the Universe. This distorts $E$-mode patterns into a combination of curl-free $E$ pattern and gradient-free $B$ patterns. The gravitational lensing signal peaks at arcminute scales, as we see in Figure \ref{fig:bk_vs_world}. It is non-Gaussian, and can be disentangled and subtracted from the primordial signal using higher order statistics.
Other sources of polarized contamination are astrophysical foregrounds, especially diffuse galactic ones. For now, only polarized dust ---due to the alignment of dust grains in the galactic magnetic fields--- and synchrotron ---due to relativistic electrons spiraling around the magnetic lines--- have been detected. Their angular power spectra follow power laws, dominating at large scales. Their frequency dependence, distinct from that of the CMB, can be used to estimate and remove them, as we will detail in the next sections.

\section{Observing $B$-modes with the BICEP/Keck Experiment }\label{sec:BK}

\begin{figure*}[!h]
\centerline{\includegraphics[trim={0 0 0 2cm},width=0.68\linewidth]{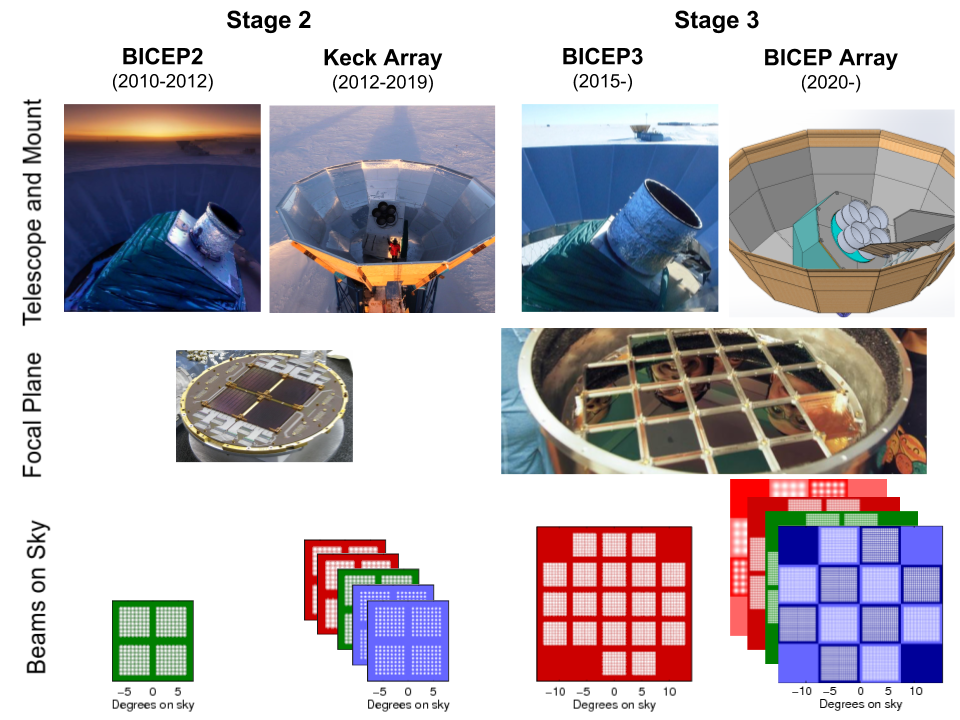}}
\caption[]{Illustration of the staged evolution of the \bk\ program. \textit{From top to bottom}, we show a picture (or a rendering for \biceptng), the focal plane of a receiver, as well as a projection on the sky of the focal plane.}
\label{fig:BKevolution1}
\end{figure*}

\bicepone, \biceptwo, \bicepthree\  and the \keckarray\  are a series of experiments located at the Amundsen-Scott South Pole Station. They have been producing the deepest degree-scale maps for the last decade. These small-aperture on-axis refracting telescopes are conceptually similar (the different stages of the experiments are shown in Figure \ref{fig:BKevolution1}). They are as compact as possible, while maintaining sufficient resolution to observe the $\ell \sim 80$ recombination bump. Their optics are cooled to 4K, improving sensitivity and allowing tight control of systematics. The entire telescopes can rotate around the boresight axis, providing a polarization modulation, and systematics cross-checks. The focal plane, cooled to $\sim 250 mK$ using a 3-stage Helium sorption refrigerator, holds an array of polarimeters, each consisting of a pair of co-located A/B orthogonal detectors. The signal is detected by voltage-biased titanium Transition Edge Sensors (TES). Data are recorded from the full array of detectors using a time-multiplexed SQUID readout system.

\begin{figure*}[!h]
\centerline{\includegraphics[trim={0 6.5cm 0 6cm},clip,width=0.72\linewidth]{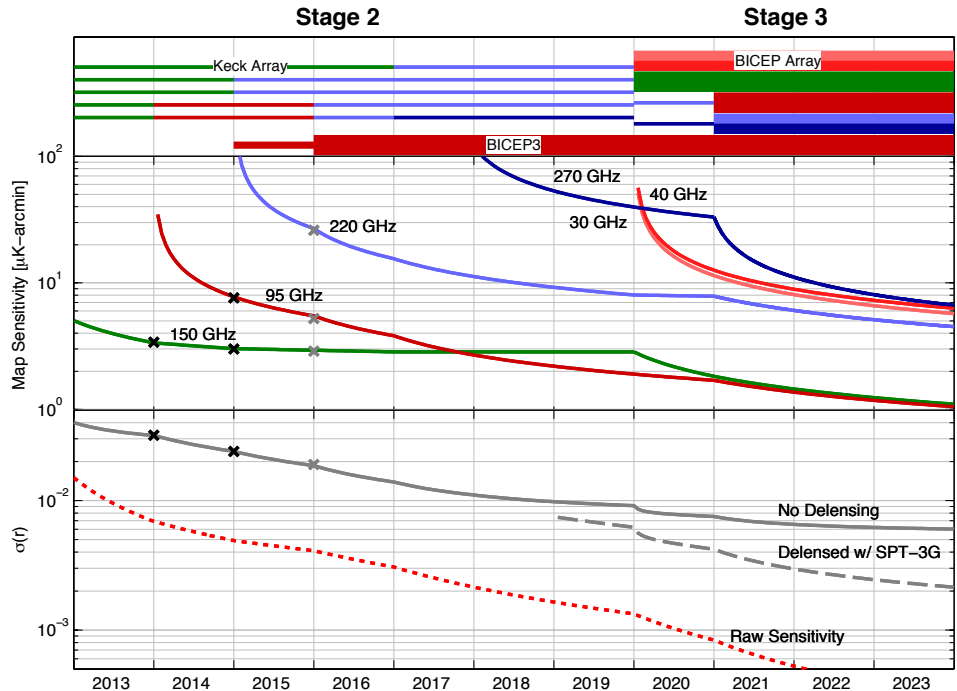}}
\caption[]{Projected sensitivity to $r$ from 2013 to 2023. \textit{Top}: Frequency distribution over the Stage 2 and Stage 3 receivers. \textit{Middle}: Time evolution of the map depths at each frequency.
\textit{Bottom}: Sensitivity to $r$ after marginalization over the seven foreground parameters, as well as the "no foreground" raw sensitivity (red). These projections involve direct scalings from published end-to-end analyses (here BK14) and hence include all real-world inefficiencies (detector yield, detector performance, weather, observing efficiency as well as incomplete mode coverage due to sky coverage, scan strategy, beam smoothing, and filtering in data analysis). The same forecasting pipeline was used for CMB Stage 4 (S4), see V. Buza's talk. Crosses represent achieved sensitivities (BKP\cite{2015BKP}, BK14\cite{2016BK14} and upcoming BK15). The dashed gray line shows the expected improvement if we delens the field using \sptthreeg\ data (from an effective residual lensing amplitude of $A_{\rm L}$=0.56 in 2018 to 0.23 in 2023, see W. L. K. Wu's talk).}
\label{fig:BKevolution2}
\end{figure*}

The high altitude and extreme dryness at the South Pole provide a highly transparent atmosphere allowing to observe in multiple frequency windows (see Figure \ref{fig:BKevolution1} and \ref{fig:BKevolution2}). \biceptwo, a 26cm diameter single-aperture telescope with 512 TES at 150 GHz (see further details about the instrument\cite{2014B2Experiment3yr}, and detectors\cite{2015BKTES}), operated from 2010 to 2012 from the Dark Sector Laboratory (DSL). \keckarray, starting in 2012, has five \biceptwo-like individual telescope receivers, operating at different frequencies (95 to 270 GHz). It is located on the large \dasi\ mount.  In 2015, \biceptwo\ was replaced by \bicepthree, a single 55cm-diameter receiver, with an optical throughput 10 times that of \biceptwo. \bicepthree\ observes at 95 GHz with 2560 detectors.

\biceptwo\ was the first instrument to detect $B$-modes at high significance\cite{2014B2results}, which were later shown to be compatible with a dust-only origin, using \planck\ higher frequency data (hereafter BKP\cite{2015BKP}).
At the time of this writing, the most recent published \bk\ paper includes all data up to 2014 (hereafter BK14\cite{2016BK14}), including data at 95 GHz for the first time. Here we discuss these results, as well as preliminary results from the BK15 analysis, doubling the amount of data at 95 GHz and including for the first time data at 220 GHz.

\begin{figure*}[!t]
\centerline{\includegraphics[trim={0 0.1cm 0 1cm},width=0.9\linewidth]{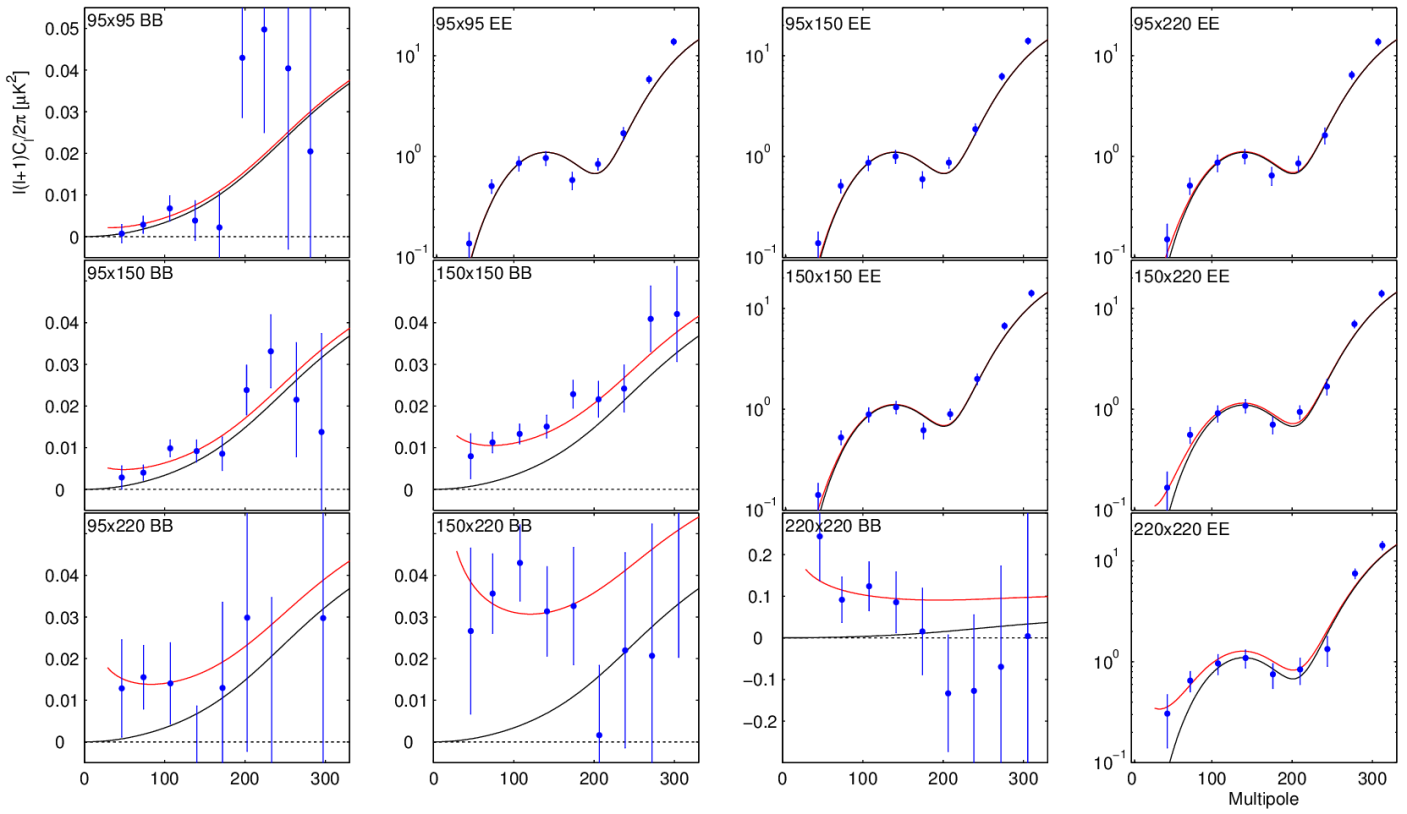}}
\caption[]{BK15 $EE$ and $BB$ cross-spectra at 95, 150 and 220 GHz. Blue points with error bars are the measured bandpowers. Black lines show the lensed-$\Lambda$CDM expectation values, and red lines show the lensed-$\Lambda$CDM+dust case for the BK14 $BB$-only baseline results ($r$ = 0, $\Ad$ = 4.3 $\mu K^2$, $\beta_d$ = 1.6, $\alpha_d$ = -0.4). The error bars are scaled to this model. The fact that the BK14 model is in good agreement with $EE$ and 220 GHz data is a validation of the model's consistency. }
\label{fig:BK15spectra}
\end{figure*}

The strategy of \bk\ is to observe one of the cleanest 1\% patches of sky continuously available from South Pole. The telescopes point at a constant elevation for 50 minutes, scan back and forth in azimuth across $\sim 60^{\circ}$ at $2.8^{\circ}/s$, and then step by $0.25^{\circ}$ in elevation. 
Immediately before and after every 50-minute scan-set, we perform an elevation nod, where the telescope scans in elevation. The resulting small change in atmospheric loading is used for relative calibration of the detectors. 

Time ordered data (TOD) go through a deglitching and relative calibration process. We then compute pair-sum and pair-difference time-streams from each A/B detector pair. While the pair-sums contain all the signal, pair-differences are only sensitive to polarized signal. A third-order polynomial filtering is applied to remove residual $1/f$ noise, which is weak in the pair-difference, since the atmosphere is largely un-polarized. We also remove signals that are fixed with respect to the ground. We perform a series of data-quality selections, and then bin this data into Temperature and Stokes Q and U maps. We transform these maps into power spectra using the matrix-based method described in our dedicated paper\cite{2016BKMatrix} to avoid $E$ to $B$ mixing. Before we look at our real power spectra, all the datasets go through a careful exercise of "jackknife" checks, where data is split into several sub-datasets designed to reveal possible systematic effects. The $BB$ and $EE$ spectra from the BK15 dataset are shown in Figure \ref{fig:BK15spectra}. The new 220 GHz data is in notably good agreement with the best-fit model from the BK14.

\section{Update on the Analysis and Latest Constraints.}\label{sec:analysis}
In order to constrain $r$ while marginalizing over foreground emissions, we make use of external publicly available data ---\planck\ PR2 polarized maps: LFI: 30, 44 and 70 GHz, and HFI: 100, 143, 217 and 353 GHz and \wmap\ nine year release: K:23 and Ka:33 GHz---, and calculate all the polarized auto- and cross-power spectra. We then compute the likelihood, with the Hamimeche-Lewis (HL) approximation\cite{2008PhRvD..77j3013H}, using a model that includes CMB from lensed-$\Lambda$CDM, and PGW of amplitude $r$, as well as dust and synchrotron emission. The $BB$ cross- or auto-power spectrum from galactic dust and synchrotron between a map A and B is defined in Table \ref{Table:8Dpar}. 

\begin{table}[t]
\begin{center}
\caption[.]{Likelihood parametrization. Our model for the foreground spectra is given by
$
\mathcal{D}_{\ell,BB}^{A \times B} =\Ad f_\mathrm{d}^A f_\mathrm{d}^B \left( \frac{\ell}{80} \right)^{\ad} +  \As f_\mathrm{s}^A f_\mathrm{s}^B \left( \frac{\ell}{80} \right)^{\as} + \epsilon \sqrt{\Ad\As} (f_\mathrm{d}^A f_\mathrm{s}^B + f_\mathrm{s}^A f_\mathrm{d}^B) \left( \frac{\ell}{80} \right)^{(\ad + \as) / 2} , 
$
 where $f_\mathrm{d}$ and $f_\mathrm{s}$ are frequency scalings that take into account the spectral energy distribution (SED) of foregrounds integrated in the bandpass of a given frequency; for synchrotron the SED is a power law  $\left( \frac{\nu}{\nu_{0,s}}\right)^{\beta_s}$, and for dust a modified blackbody (MBB), $ \left( \frac{\nu}{\nu_{0,d}}\right)^{\beta_d}\frac{B_\nu(T_d)}{B_{\nu_{0,d}}(T_d)}$, with $T_d=19.6K$. Flat priors are reported as bracketed [minimum, maximum], whereas Gaussian priors with mean $\mu$ and standard deviation $\sigma$ are reported as $\mathcal{N}(\mu,\sigma)$, together with the published source. The only change in BK15 relative BK14 is the flat prior on $\epsilon$, going from [0,1] to [-1,1]. The eight parameters in the table define the baseline analysis.}
\label{Table:8Dpar}

\begin{tabular}[b]{|c|l|c|}
\hline
  \textbf{Parameter}             & \textbf{Description}      & \textbf{Prior} \\
\hline
$r$         & Tensor-to-scalar power ratio, at a pivot scale of 0.05 Mpc$^{-1}$    &  [0,0.5] \\
$\Ad$	& Dust amplitude in $\mu K^2$, at $\ell=80$ and $\nu_{0,d}=353$ GHz & [0,15] $\mu K^2$\\
$\As $     & Synchrotron amplitude in $\mu K^2$, at $\ell=80$ and $\nu_{0,s}=23$ GHz & [0,50] $\mu K^2 $\\
$\beta_d$       & Dust spectral index	& $\mathcal{N}(1.59,0.11)$\cite{2015A&A...576A.107P}\\
$\beta_s$      & Synchrotron spectral index	&  $\mathcal{N}(-3.1,0.3)$\cite{2014ApJ...790..104F} \\
$\ad$ & Dust angular power spectrum power law index & [-1,0]\\
$\as$     & Synchrotron angular power spectrum power law index	  &  [-1,0] \\
$\epsilon$ & 	Dust-Synchrotron correlation &  \textbf{[-1,1]} \\
\hline

\end{tabular}
\end{center}

\end{table}

\begin{figure}[!h]

\centerline{\includegraphics[trim={0 0.cm 0 0cm},width=0.65\linewidth]{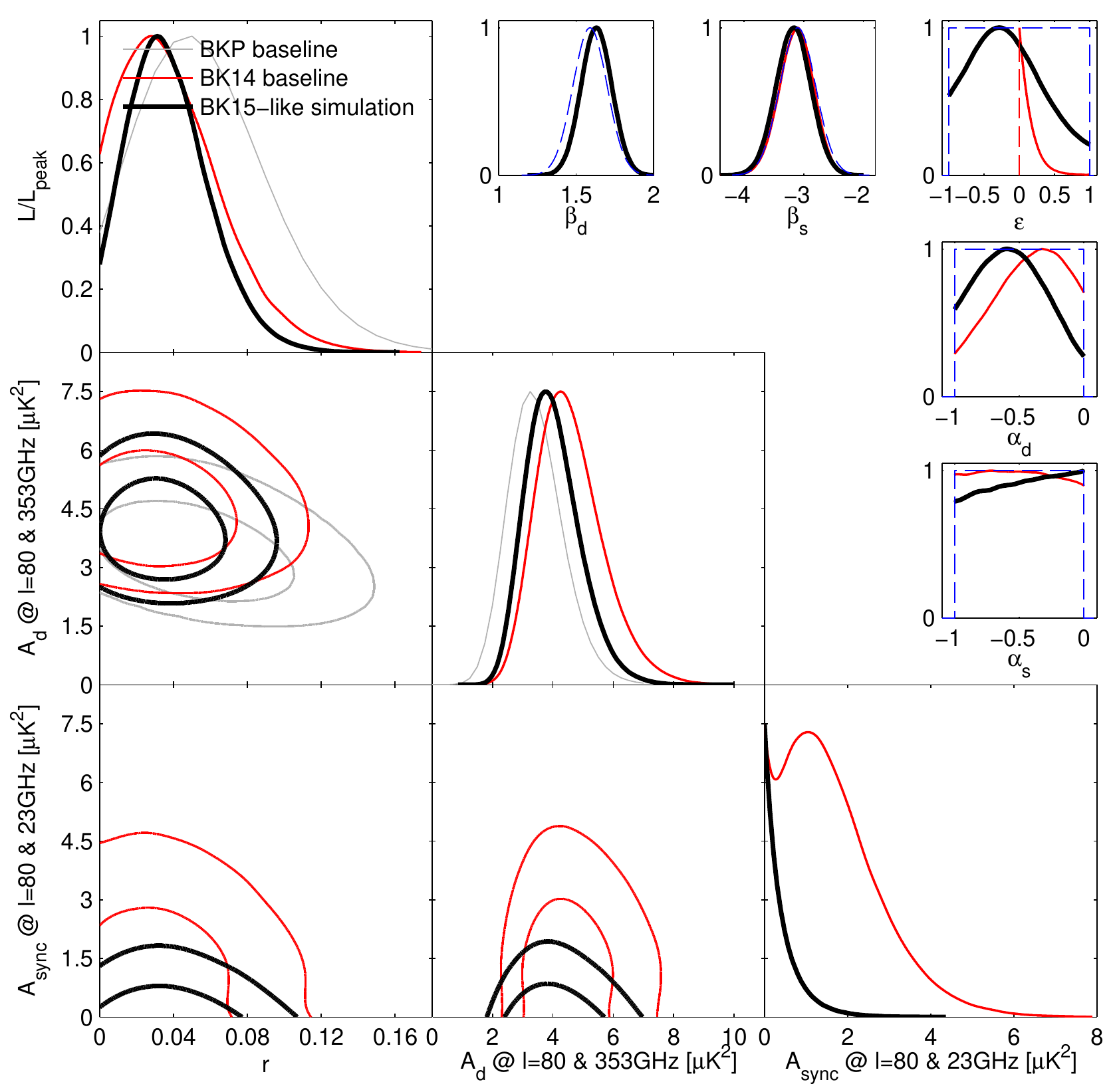}}

\caption[]{"Baseline" 8D likelihood curves for a BK15-like simulation. Our simulations contain Gaussian dust maps, with MBB SED, lensed-$\Lambda$CDM CMB, and BK15 frequency coverage and noise levels. This simulation was chosen because of its similarity to BK14 for the peak position of $r$ and $\Ad$. . The triangle plot shows 1D and 2D marginalized likelihoods for $r$, $\Ad$ and $\As$, and the smaller plots show the 1D ones for the remaining parameters. The imposed priors are shown in dashed blue (except for the $\epsilon$ prior from BK14 in dashed red. In gray we show the BKP results, for a 2D likelihood, in red the BK14 results, and in black the simulation. }
\label{fig:BK15sims_like1}
\end{figure}

\begin{figure}[!h]

\centerline{\includegraphics[trim={0 0.cm 0 0cm},width=0.99\linewidth]{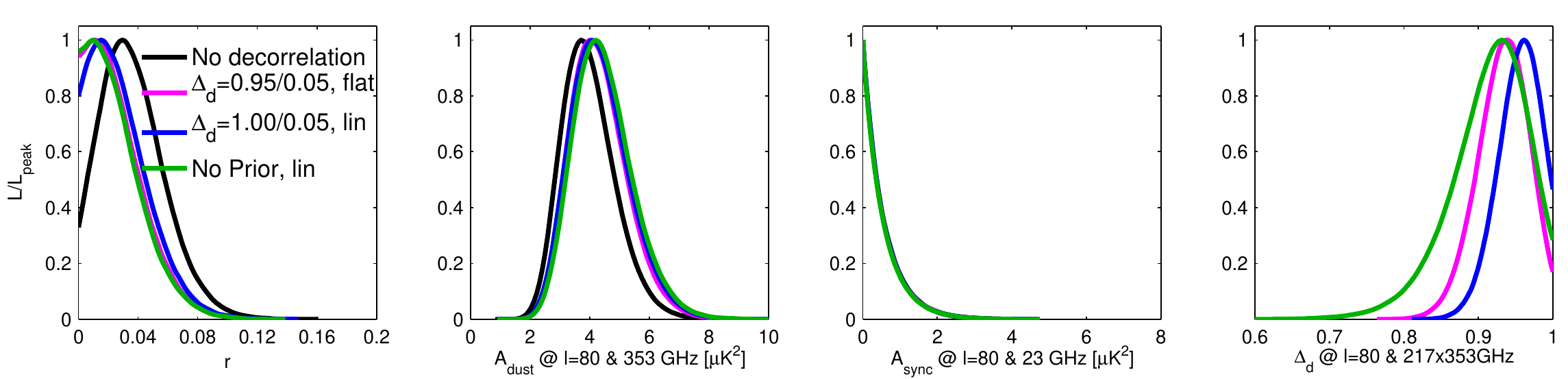}}
\caption[]{Likelihood results (simulation) when allowing for dust decorrelation.}

\label{fig:BK15sims_like2}
\end{figure}

\begin{figure}[!h]

\centerline{\includegraphics[trim={0 0.cm 0 0cm},width=0.8\linewidth]{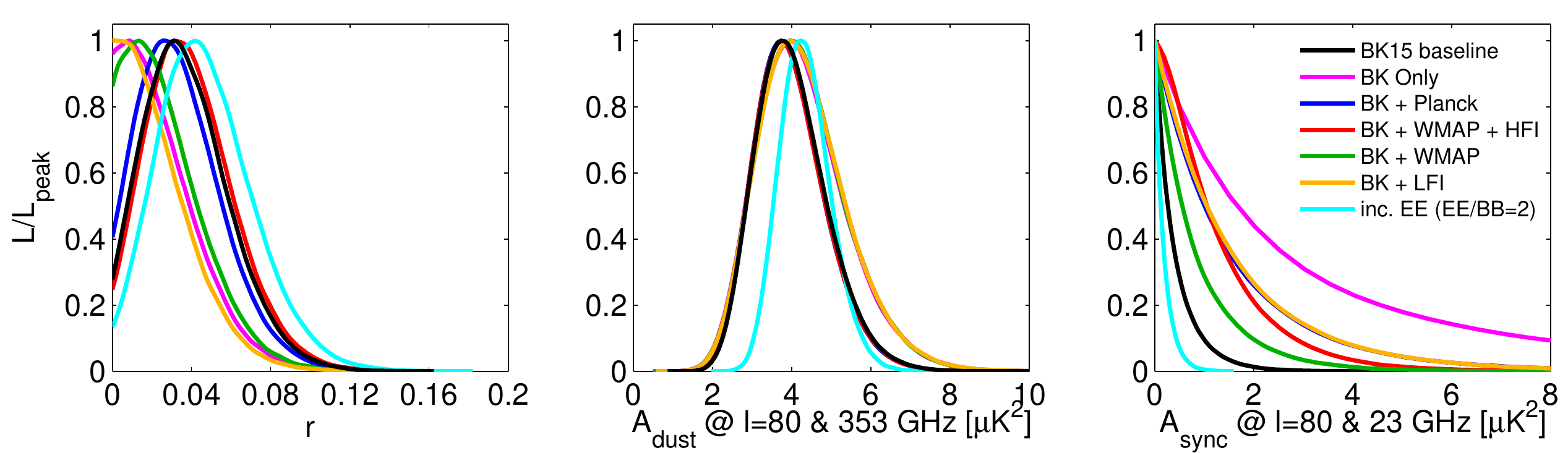}}
\caption[]{Likelihood results (simulation) when varying the data set selection. }

\label{fig:BK15sims_like3}
\end{figure}

In the BK14 and BK15 analysis, we use 9 bandpowers between $\ell=20$ and 330. We feed in all the spectra, as well as their full covariance matrix ---derived from signal and noise simulations--- into \texttt{COSMOMC}\cite{2002PhRvD..66j3511L}. The baseline analysis results for BK14 are shown in Figure \ref{fig:BK15sims_like1}, along with a "BK15-like" simulation. For the first time in BK14, we obtained better constraints from B-modes only ($r_{0.05}<0.09$ at 95\% confidence) than from the best non-B-mode results ($r_{0.05}<0.12$\cite{2016A&A...594A..13P}). Note that the constraint would be even better if the posterior peaked at zero. BK14 has a strong detection of dust but no detection of synchrotron. 
Going from BKP to BK14, we see a decrease in the width of the $r$ posterior (from $\sigma(r)\sim0.034$ to 0.025) as well as a reduction of the degeneracy between $r$ and $\Ad$ . These are the tightest constraints to date on PGW. With BK15, we expect a further tightening of the contours, to $\sigma(r)\sim0.019$.

With more data, we can also perform other tests of the robustness of the results . In the upcoming BK15 paper,  we test how the likelihood varies with data set selection, analysis choices (HL fiducial model, priors on foreground SED etc.), or lensing amplitude. We also study how the introduction of dust decorrelation affects the likelihood analysis. In past analyses, we assumed that the galactic dust foreground could be approximated as a given emission template that scales in frequency according to a given SED. In reality, different clouds of dust have different temperatures (hence different SEDs) and different mean polarization orientations. Due to these variations of the dust SED and polarization angles,  dust emission will not be fully correlated at different frequencies. \planck\ first reported a detection of such a decorrelation\cite{2017A&A...599A..51P}, but later analyses showed that the level is in fact below the current instrumental noise\cite{2018arXiv180104945P,2018PhRvD..97d3522S}. In the BK15 analysis we search for evidence of dust decorrelation. In Figure \ref{fig:BK15sims_like2}, we show how the $r$, $\Ad$, $\As$ and $\ddp$ posteriors vary for the same BK15-like simulation, given different priors and $\ell$-dependence of the decorrelation. Even though there is no decorrelation ($\ddp$=1) in the input model of the simulation, the $\ddp$ posterior doesn't peak at 1, and more importantly the peak of the $r$ posterior gets reduced by roughly $1\sigma$ when decorrelation is marginalized over. The fact that we impose physical bounds on these correlated parameters implies that including $\ddp$ can bias low the estimate of $r$. We studied this effect in detail in the BK15 paper and decided not to include dust decorrelation in the baseline analysis. 

High frequency \keck\ observations increase the constraining power on dust. We analyzed our model with BK only data, as well as with subsets of \planck\ and/or \wmap. Figure \ref{fig:BK15sims_like3} shows these analysis variations on the selected BK15-like simulation. We observe shifts in the $r$ posterior and in the synchrotron constraints, all consistent with expected statistics.

\section{Future plans.}\label{sec:future}
In 2016 and 2017, we have accumulated deep observations at 220 GHz using \keck\ (and even some 270 GHz data for the first time) as well as 95 GHz using \bicepthree. These will allow an even better mitigation of foregrounds. We forecast $\sigma(r) \lessapprox 0.01$, using performance-based estimates.

\biceptng\ (BA) is the Stage 3 modular telescope that will replace \keckarray\ on a new mount. It will have \bicepthree-like receivers with the modularity of \keck, observing at 30, 40, 95, 150, 220 and 270 GHz (see Figure \ref{fig:BKevolution1} and \ref{fig:BKevolution2}). It will be deployed during the Austral summer 2019-2020. Microwave SQUID multiplexers will allow the 220/270 GHz receiver to have $\sim 22000$ detectors. It will surpass the dust sensitivity of \planck\ 353 GHz channel within a few days. The 30/40 GHz receiver will also rapidly improve the measurements of polarized synchrotron. 

We also plan a wide survey of $\sim 20\%$ of the sky, which will improve over \planck\ S/N within a few months. This will be useful in order to look for the cleanest patches of sky in the southern hemisphere. These other regions could be used to confirm a cosmological signal if observed in the original patch. We will also be able to test the general validity of the foreground models with more data. It will also of course be interesting for studies of galactic foregrounds, especially the anomalous microwave emission.

With the sensitivity achieved by BA, it will become even more necessary to de-lens the $B$-mode map, i.e. remove at the map level the signal due to gravitational lensing. This will be possible thanks to high resolution $E$-modes observations from \sptthreeg. Using higher order statistics on $E$ and $B$ maps, the lensing B-modes can be reconstructed and taken into account in the data analysis.  The lensing template can be added as an extra component to the likelihood analysis. While having promising results on simulations using an external tracer for the template, more advanced techniques will be necessary for optimal delensing. A proof-of-concept \bk\ x \sptpol\ delensing analysis is currently in progress (see W. L. K. Wu's talk).

\cmbsfour\ (S4) is an effort of current teams --- \bk, \spt, \act, \polarbear, Simons Observatory --- to build multiple telescopes with a shared design with a total of over half a million detectors. For the ``Inflation'' science goal, S4 plans to have a mix of small aperture telescopes targeting the recombination bump, and large aperture telescopes used for delensing. The combined analysis of \biceptng\ and \sptthreeg\ will be a major step towards this goal.

\section*{Acknowledgments}
The team gives a special acknowledgment to its heroic multi-season winter-overs Robert Schwarz, Steffen Richter, and Hans Boenish. B.R. thanks the organizers of the Rencontres de Moriond once again for a great conference.

\begin{multicols}{2}

\end{multicols}


\begin{thebibliography}{99}
\footnotesize
\bibitem{2016A&A...594A..13P} \planck\ Collaboration, \ 2016, \aap, 594, A13 
\bibitem{1981PhRvD..23..347G} Guth, A.~H.\ 1981, \PRD, 23, 347 
\bibitem{1981ZhPmR..33..549M} Mukhanov, V.~F., \& Chibisov, G.~V.\ 1981, {\em Pisma Zh.Eksp.Teor.Fiz}, 33, 549 
\bibitem{1997EBSeljak} Seljak, U., \& Zaldarriaga, M.\ 1997, \PRL, 78, 2054 
\bibitem{1997EBKamionkowski} Kamionkowski, M., Kosowsky, A., \& Stebbins, A.\ 1997, \PRL, 78, 2058 
\bibitem{2014B2Experiment3yr} \biceptwo\ Collaboration, \ 2014, \APJ, 792, 62 
\bibitem{2015BKTES} \biceptwo, \keckarray\ \& \spider\ Collaborations, \ 2015, \APJ, 812, 176 
\bibitem{2014B2results} \biceptwo\ Collaboration, \ 2014, \PRL, 112, 241101 
\bibitem{2015BKP} \biceptwo, \keckarray\ \& \planck\ Collaborations, \ 2015, \PRL, 114, 101301 
\bibitem{2016BK14} \biceptwo\ \& \keckarray\ Collaborations, \ 2016, \PRL, 116, 031302 
\bibitem{2016BKMatrix} \biceptwo\ \& \keckarray\ Collaborations, \ 2016, \APJ, 825, 66 
\bibitem{2008PhRvD..77j3013H} Hamimeche, S., \& Lewis, A.\ 2008, \PRD, 77, 103013 
\bibitem{2015A&A...576A.107P} Planck Collaboration, \ 2015, \aap, 576, A107 
\bibitem{2014ApJ...790..104F} Fuskeland, U., Wehus, I.~K., Eriksen, H.~K., \& N{\ae}ss, S.~K.\ 2014, \APJ, 790, 104 
\bibitem{2002PhRvD..66j3511L} Lewis, A., \& Bridle, S.\ 2002, \PRD, 66, 103511 
\bibitem{2017A&A...599A..51P} \planck\ Collaboration,\ 2017, \aap, 599, A51 
\bibitem{2018arXiv180104945P} \planck\ Collaboration, \ 2018, arXiv:1801.04945 
\bibitem{2018PhRvD..97d3522S} Sheehy, C., \& Slosar, A.\ 2018, \PRD, 97, 043522 
\end{thebibliography}
\end{document}